\documentclass[aps,pra,amssymb,amsmath,eqsecnum,nofootinbib]{revtex4}
\usepackage{graphicx}

\newtheorem{remark}{Remark}[section]

\begin{document}
\title{Phase operator of the quantum supersymmetric harmonic oscillator}
\author{Gavriel Segre}
\homepage{http:\\www.gavrielsegre.com}
\begin{abstract}
After a brief introduction recalling how, in the limit in which
the mass and the electric charge of the electron and the positron
tend to zero, Quantum Electrodynamics reduces to a collection of
uncoupled  quantum supersymmetric harmonic oscillators, the phase
operator of the quantum fermionic harmonic oscillator and of the
quantum supersymmetric harmonic oscillator are introduced and
their properties analyzed.

It is then shown that the phase operator of a supersymmetric
harmonic oscillator is a Goldstone operator at any strictly
positive temperature (finite or infinite).
\end{abstract}
\maketitle
\newpage
\tableofcontents
\newpage
\section{Acknowledgements}

I would like to thank strongly Vittorio de Alfaro for his
friendship and his moral support, without which I would have
already given up.

Then I would like to thank strongly Jack Morava for many precious
teachings.

Finally I would like to thank strongly Andrei Khrennikov and the
whole team at the International Center of Mathematical Modelling
in Physics and Cognitive Sciences of V\"{a}xj\"{o} for their very
generous informatics' support.

Of course nobody among the mentioned people has responsibilities
as to any (eventual) error contained in these pages.
\newpage
\section{Introduction}

A free field theory is equivalent to a collection of harmonic
oscillators.

The quantization of such a field theory simply reduces to the
quantization of these oscillators.

The bosonic or fermionic nature of the involved field theory
determines whether the resulting quantum harmonic oscillators are
bosonic or fermionic.

A longstanding issue in the framework of Quantum Optics concerns
the definition of the phase operator for a bosonic harmonic
oscillator and the resulting phase properties of the quantum
electromagnetic field (see for instance
\cite{Freyberger-Vogel-Schleich-O-Connell-06}, the section 2.8
"Phase Properties of the Field" of \cite{Walls-Milburn-94}, the $
4^{th} $ chapter "Phase Operator" of \cite{Bendjaballah-95} as
well as \cite{Barnett-Vaccaro-07a}, \cite{Barnett-Vaccaro-07b},
\cite{Barnett-Vaccaro-07c}, \cite{Barnett-Vaccaro-07d},
\cite{Barnett-Vaccaro-07e} , \cite{Barnett-Vaccaro-07f}).

Such an issue is deeply linked with the issue concerning the
impossibility of  defining a time operator in Quantum Mechanics
\cite{Barnett-Vaccaro-07g} and is, hence, deeply linked with the
previous research about time we performed in \cite{Segre-07b} and
in \cite{Segre-07c}.

Though  considering the recent wonderful book edited by Stephen M.
Barnett and John A. Vaccaro with no doubt the best existing
rescource concerning the issue of defining the phase operator for
a quantum bosonic oscillator, we must confess that we don't agree
with the viewpoint of the authors since we think that the
\emph{Pegg-Barnett operator}, introduced by D.T. Pegg and S.M.
Barnett in  their 1988-1989's papers (now available as
\cite{Pegg-Barnett-07a}, \cite{Pegg-Barnett-07b},
\cite{Pegg-Barnett-07c}, \cite{Pegg-Barnett-07d}) recovers the
self-adjointness lacking to the \emph{Susskind-Glogover operator}
(introduced by Leonard Susskind and Jonathan Glogover in their
1964's paper now available as \cite{Susskind-Glogover-07} and
supported by R. Loudon in the $ 7^{th} $ chapter of the first
1973's edition of his manual of Quantum Optics now available as
\cite{Loudon-07}) only at the prize of a considerable decrease in
the formal elegance and beauty.

Furthermore nowadays it has become generally accepted to consider
as the set of the physical observables of a (closed) quantum
system something bigger than the set of all the self-adjoint
operators on the system's Hilbert space $ \mathcal{H}_{system} $
commuting with all the \emph{superselection charges}.

For instance \emph{unsharp observables} (i.e.
\emph{positive-operator valued measures} that are not
\emph{projection valued measures}  and hence are not equivalent,
via the \emph{Spectral Theorem} \cite{Reed-Simon-80}, to a
self-adjoint operator over $ \mathcal{H}_{system} $ \footnote{Let
us recall that, by Naimark's Theorem, a \emph{positive-operator
valued measure} over $ \mathcal{H}_{system} $ may be seen as a
projection valued measure on a suitably enlarged Hilbert space,
though this fact, together with the acceptance of \emph{unsharp
observables} (and of \emph{non-projective measurements})  doesn't
solve the Measurement Problem of Quantum Mechanics contrary to
what it is sometimes believed \cite{Busch-Grabowski-Lahti-95}.})
are nowadays generally accepted.

For this reasons we believe that the correct  \emph{phase
operator} for a bosonic harmonic oscillator is the
\emph{Susskind-Glogover operator} to which will refer from here
and beyond as the \emph{bosonic phase operator.}

\bigskip

Let us now remark that the following  analog problems, i.e.:
\begin{enumerate}
    \item to define the phase operator for the quantum
fermionic harmonic oscillator
    \item to define the phase operator for the quantum supersymmetric
harmonic oscillator
\end{enumerate}

haven't been, at least up to our knowledge, investigated yet.

This is curious since, in the framework of Relativistic Quantum
Mechanics,  we are used nowadays to think that we know everything
concerning Quantum Electrodynamics \cite{Mandl-Shaw-93},
\cite{Itzykson-Zuber-05}, \cite{Weinberg-95}, \cite{Weinberg-96}.

Anyway it is sufficient to consider the limit in which the
electric charge and the mass of the electron and of the positron
tend to zero of the QED's quantum field theory having lagrangian
density \footnote{adopting the usual notation where:
\begin{equation}
    F_{\mu \nu} \; := \; \partial_{\mu} A_{\nu} - \partial_{\nu} A_{\mu}
\end{equation}
 $ \{ \gamma^{0} ,  \gamma^{1} , \gamma^{2} , \gamma^{3} \} $ are Dirac matrices, i.e.  $
4 \times 4 $  matrices satisfying the condition:
\begin{equation}
    \{ \gamma^{\mu} , \gamma^{\nu} \} \; = \; 2 \eta^{\mu \nu}
\end{equation}
 where $ ( \mathbb{R}^{4} , \eta := \eta_{\mu \nu} d x^{\mu} \otimes d
x^{\nu} ) $ is the Minkowski spacetime with $ \eta_{\mu \nu} \; :=
\; diag( 1 , -1 , -1 , -1)$, where $ \psi $ is a 4-component
spinor, where $ \bar{\psi} := \psi^{\dag} \gamma^{0} $, where
Einstein's convention  of sum over repeated indices is assumed and
where indices are raised and lowered by contraction with the
Minkowskian metric tensor.}:
\begin{equation}
    \mathcal{L}_{QED} \; := \; - \frac{1}{4} F_{\mu \nu} F^{\mu
    \nu} + \bar{\psi} ( i \gamma^{\mu} D_{\mu} - m ) \psi
\end{equation}
\begin{equation}
    D_{\mu} \; := \; \partial_{\mu} - i e A_{\mu}
\end{equation}  having  quantum hamiltonian
\footnote{imposing for simplicity periodic boundary conditions on
the walls of a cube of side L.}:
\begin{equation}
    H \; =  \;H_{B} + H_{F}
\end{equation}
where:
\begin{equation}
    H_{B} \; := \;  \sum_{\vec{k} \in \frac{ 2 \pi }{L} \mathbb{Z}^{3}}
    \sum_{r=0}^{3}  \omega_{B;\vec{k}}  N_{B ; \vec{k},r}
\end{equation}
\begin{equation}
    \omega_{B; \vec{k}} \; := \; | \vec{k} |
\end{equation}
\begin{equation}
    N_{B; \vec{k} , r } \; := \; \zeta_{r} a_{B; \vec{k},r}^{\dag}
    a_{B; \vec{k},r}
\end{equation}
\begin{equation}
    \zeta_{r} \; := \; \left\{%
\begin{array}{ll}
    -1, & \hbox{if $ r= 0$} \\
    1, & \hbox{if $ r \in \{ 1,2,3 \}$} \\
\end{array}%
\right.
\end{equation}
\begin{equation}
    [ a_{B; \vec{k},r} , a_{B; \vec{k}',s}^{\dag} ] \; = \;
    \delta_{rs} \delta_{\vec{k} , \vec{k}'}
\end{equation}
\begin{equation}
    [ a_{B; \vec{k},r} , a_{B; \vec{k}',s} ] \; = \; [ a_{B; \vec{k},r}^{\dag} , a_{B; \vec{k}',s}^{\dag}
    ] \; = \; 0
\end{equation}
\begin{equation}
    H_{F} \; := \; \sum_{\vec{k} \in \frac{ 2 \pi }{L} \mathbb{Z}^{3}}
    \sum_{r \in \{ 1 , 2 \}}  \omega_{F;\vec{k}} (  N_{F, + ;
    \vec{k},r} + N_{F, - ; \vec{k},r}  )
\end{equation}
\begin{equation}
  \omega_{F;\vec{k}} \; := \; \sqrt{ \vec{k}^{2} + m^{2} }
\end{equation}
\begin{equation}
    N_{F, + ; \vec{k},r} \; := \; a_{F,+; \vec{k},r}^{\dag} a_{F,+; \vec{k},r}
\end{equation}
\begin{equation}
    N_{F, - ; \vec{k},r} \; := \; a_{F,-; \vec{k},r}^{\dag} a_{F,-; \vec{k},r}
\end{equation}
\begin{equation}
    \{ a_{F,+; \vec{k},r} ,  a_{F,+; \vec{k}',s}^{\dag} \} \; = \;
    \delta_{\vec{k},\vec{k}'} \delta_{r,s}
\end{equation}
\begin{equation}
    \{ a_{F,-; \vec{k},r} ,  a_{F,-; \vec{k}',s}^{\dag} \} \; = \;
    \delta_{\vec{k},\vec{k}'} \delta_{r,s}
\end{equation}
\begin{equation}
    \{  a_{F,+; \vec{k},r} , a_{F,+; \vec{k}',s} \} \; = \; \{  a_{F,+; \vec{k},r}^{\dag} , a_{F,+; \vec{k}',s}^{\dag}
    \} \; = \;  0
\end{equation}
\begin{equation}
    \{  a_{F,-; \vec{k},r} , a_{F,-; \vec{k}',s} \} \; = \; \{  a_{F, -; \vec{k},r}^{\dag} , a_{F,-;
    \vec{k}',s}^{\dag} \} \; = \; 0
\end{equation}
\begin{equation}
    \{  a_{F,+; \vec{k},r} , a_{F,-; \vec{k}',s} \} \; = \; \{  a_{F,+; \vec{k},r} ,  a_{F,-;
    \vec{k}',s}^{\dag} \} \; = \; \{  a_{F,+; \vec{k},r}^{\dag} ,  a_{F,-;
    \vec{k}',s} \} \; = \; \{ a_{F,+; \vec{k},r}^{\dag} , a_{F,-;
    \vec{k}',s}^{\dag} \} \; = \; 0
\end{equation}
to obtain a system of uncoupled quantum supersymmetric oscillators
whose possible physical states are the rays of the $
\mathbb{Z}_{2}$-graded Hilbert space $ \mathcal{H}_{B}^{physical}
\otimes \mathcal{H}_{F} $, where $ \mathcal{H}_{B}^{physical} $ is
the subspace of $ \mathcal{H}_{B} $ obtained imposing the Lorentz
gauge condition $ \partial_{\mu} A^{\mu} = 0 $ in the Gupta-Bleuer
form:
\begin{equation}
  \mathcal{H}_{B}^{physical} \; := \; \{ | \psi > \in
  \mathcal{H}_{B} \,  : \, ( a_{B; \vec{k},3} - a_{B; \vec{k},0}) | \psi
  > \; = \; 0 \; \; \forall \vec{k} \in \frac{2 \pi  }{L}
  \mathbb{Z}^{3} \}
\end{equation}
as it appears evident as soon as one expresses the restriction of
the quantum hamiltonian H to $ \mathcal{H}_{B}^{physical} \otimes
\mathcal{H}_{F} $ as:
\begin{equation}
    H |_{\mathcal{H}_{B}^{physical}
\otimes \mathcal{H}_{F}} \; = \; \sum_{\vec{k} \in \frac{ 2 \pi
}{L} \mathbb{Z}^{3}} \sum_{r \in \{1,2  \} } | \vec{k} | (  N_{B ;
\vec{k},r} + N_{F ;\vec{k},r } )
\end{equation}
\begin{equation}
   N_{F ;\vec{k},r } \; := \; N_{F, + ; \vec{k},r} +  N_{F,- ;
   \vec{k},r} \; \; \vec{k} \in \frac{ 2 \pi
}{L} \mathbb{Z}^{3} , r \{ 1 , 2 \}
\end{equation}

\smallskip

Curiously the phase properties of such a collection of uncoupled
quantum supersymmetric harmonic oscillators have not been
investigated yet.

\smallskip

In this paper we  introduce the \emph{fermionic phase operator},
i.e. the \emph{phase operator} for a \emph{fermionic harmonic
oscillator},  and the \emph{supersymmetric phase operator}, i.e.
the \emph{phase operator} for a \emph{supersymmetric harmonic
oscillator}.

Furthermore we show that the \emph{supersymmetric phase operator}
is a Goldstone operator at any strictly positive temperature.
\newpage
\part{Theory at zero temperature.}
\section{Phase operator of the quantum bosonic oscillator}
\label{sec:Phase operator of the quantum bosonic oscillator} Let
us consider a quantum bosonic oscillator having, hence,
hamiltonian:
\begin{equation} \label{eq:hamiltonian of the bosonic oscillator}
    H_{B} \; := \; \omega N_{B}
\end{equation}
where $ a_{B} $ and $ a_{B}^{\dag} $ are respectively the
annihilation and the creation operators:
\begin{equation} \label{eq:1th commutation relation for the bosonic creation and annihilation operators}
    [  a_{B} ,  a_{B} ] \; = \; [  a_{B}^{\dag} ,  a_{B}^{\dag} ]
    \; = \; 0
\end{equation}
\begin{equation} \label{eq:2th commutation relation for the bosonic creation and annihilation operators}
      [  a_{B} , a_{B}^{\dag} ] \; = \; 1
\end{equation}
and where $ N_{B} $ is the number operator:
\begin{equation}
   N_{B} \; := \; a_{B}^{\dag} a_{B}
\end{equation}
The equation \ref{eq:1th commutation relation for the bosonic
creation and annihilation operators} and the equation \ref{eq:2th
commutation relation for the bosonic creation and annihilation
operators} imply that:
\begin{equation}\label{eq:commutation relation between bosonic number and annihilation operator}
    [ N_{B} , a_{B} ] \; = \; - a_{B}
\end{equation}
\begin{equation}\label{eq:commutation relation between bosonic number and creation operator}
     [ N_{B} , a_{B}^{\dag} ] \; = \;  a_{B}^{\dag}
\end{equation}
 that imply that:
\begin{equation}
    a_{B} | n > \; = \; \left\{%
\begin{array}{ll}
    0, & \hbox{if $n=0$;} \\
    \sqrt{n} | n -1 >, & \hbox{if $ n \in \mathbb{N}_{+}$.} \\
\end{array}%
\right.
\end{equation}
\begin{equation}
   a_{B}^{\dag} | n > \; = \;  \sqrt{n+1} | n +1 > \;
  \; \forall n \in \mathbb{N}
\end{equation}
\begin{equation}
  | n > \; = \; \frac{( a_{B}^{\dag} )^{n}}{ \sqrt{n !} } | 0 > \;
  \; \forall n \in \mathbb{N}
\end{equation}
\begin{equation}
    N_{B} | n > \; = \; n | n > \; \; \forall n \in \mathbb{N}
\end{equation}
\begin{equation}
    H_{B} | n > \; = \; E_{B}(n) | n >  \; \; \forall n \in \mathbb{N}
\end{equation}
where:
\begin{equation} \label{eq:energy levels of the bosonic oscillator}
   E_{B}(n) \; := \; \omega \, n \; \; n \in \mathbb{N}
\end{equation}

\bigskip

 Let us now  introduce the \emph{bosonic angle states}:
\begin{equation} \label{eq:bosonic angle states}
    | \theta > \; := \; \frac{1}{\sqrt{ 2 \pi }}
    \sum_{n=0}^{\infty} \exp ( i n \theta ) | n > \; \; \theta \in
    [ 0 , 2 \pi )
\end{equation}

\bigskip

\begin{remark}
\end{remark}
Let us remark that:
\begin{equation}
    < \theta_{1} | \theta_{2} > \; = \frac{1}{2 \pi} \sum_{n=0}^{\infty} \exp [ i n ( \theta_{2} - \theta_{1} ) ] \neq \; \delta (  \theta_{1} -
    \theta_{2} ) \; \; \forall \theta_{1}, \theta_{2} \in [ 0 , 2
    \pi )
\end{equation}
 though clearly:
\begin{equation}
  < \theta | \theta > \; = \; + \infty \; \; \forall \theta \in [
  0 , 2 \pi )
\end{equation}
Though not orthonormal, the bosonic angle states are complete:
\begin{equation} \label{eq:completeness of the bosonic angle states}
    \int_{0}^{2 \pi} d \theta | \theta > < \theta | \; = \; \frac{1}{2
    \pi} \sum_{n=0}^{\infty}  \sum_{m=0}^{\infty}\int_{0}^{2 \pi} d \theta \exp [ i ( n - m ) \theta
    ] | n > < m | \; = \;
     \sum_{n=0}^{\infty}  \sum_{m=0}^{\infty}
     \delta_{n,m}  | n > < m | \; = \; \sum_{n=0}^{\infty} | n > <
     n | \; = \; 1
\end{equation}
where we have used the fact that:
\begin{equation} \label{eq:useful integral}
   \int_{0}^{2 \pi} d \theta \exp [ i ( n - m ) \theta
    ] \; = \;  2 \pi \delta_{n,m} \; \; \forall n,m \in \mathbb{N}
\end{equation}

\bigskip

Let us introduce the \emph{bosonic exponential phase operator}:
\begin{equation} \label{eq:bosonic exponential phase operator}
    \exp ( i \hat{\theta} ) \; := \; \sum_{n=0}^{\infty} | n > < n
    + 1 |
\end{equation}
whose name is justified by the fact that:
\begin{multline}
     \exp ( i \hat{\theta} )  | \theta >  \; = \; \frac{1}{\sqrt{ 2 \pi } } \sum_{n=0}^{\infty}  \sum_{m=0}^{\infty} | n > < n+1 | m >  \exp ( i m \theta
     ) \; = \; \frac{1}{\sqrt{ 2 \pi } } \sum_{n=0}^{\infty}  \sum_{m=0}^{\infty} \delta_{m,n+1} \exp ( i m \theta
     )  | n >  \; = \\
     \frac{1}{\sqrt{ 2 \pi } } \sum_{n=0}^{\infty} \exp [ i ( n+1 )
     \theta ] | n >  \; = \;
  \exp ( i \theta
     ) | \theta > \; \; \forall \theta \in [ 0 , 2 \pi )
\end{multline}

\begin{remark}
\end{remark}
Let us remark that the \emph{exponential bosonic phase operator} $
\exp ( i \hat{\theta} ) $ is not unitary and, hence, the
\emph{bosonic phase operator} $ \hat{\theta} $ is not
self-adjoint.

In fact:
\begin{multline}
  exp ( i \hat{\theta} )  ( exp (  i \hat{\theta} ) )^{\dag}  \; = \; (
  \sum_{n=0}^{\infty} | n > < n+ 1| ) ( \sum_{m=0}^{\infty} | m +
  1 > < m | ) \; = \; \sum_{n=0}^{\infty} \sum_{m=0}^{\infty} | n >
  < n+1 | m +1 > < m | \ = \\ \sum_{n=0}^{\infty}
  \sum_{m=0}^{\infty} \delta_{m+1, n+1} | n > < m | \; = \; \sum_{n=0}^{\infty}
  \sum_{m=0}^{\infty} \delta_{m, n} | n > < m | \; = \; \sum_{n=0}^{\infty}  | n > < n
  | \; = \; 1
\end{multline}
but:
\begin{multline} \label{eq:unitarity failure of the bosonic exponential phase operator}
  ( exp ( i \hat{\theta} ) )^{\dag} exp (  i \hat{\theta} ) \; = \; (
  \sum_{n=0}^{\infty} | n+1 > < n| ) ( \sum_{m=0}^{\infty} | m
   > < m +1 |) \; = \; \sum_{n=0}^{\infty} \sum_{m=0}^{\infty} | n+1 >
  <n | m > < m+1 | \ = \\ \sum_{n=0}^{\infty}
  \sum_{m=0}^{\infty} \delta_{m, n} | n +1 > < m +1 | \; = \; \sum_{n=0}^{\infty}  | n +1 > <
  n+1
  | \; = \; \sum_{n=1}^{\infty}  | n  > <
  n | \; = \; \sum_{n=0}^{\infty}  | n  > <
  n | - | 0 > <0| \; = \; 1 - | 0 > < 0 |
\end{multline}

\bigskip

Given a generic normalized state:
\begin{equation}
    | \psi > \; := \; \sum_{n=0}^{\infty} c_{n} | n >
\end{equation}
\begin{equation}
  < \psi | \psi > \; = \; \sum_{n=0}^{\infty} | c_{n} |^{2} \; =
  \;1
\end{equation}
the probability that a measurement of the \emph{bosonic phase
operator} $ \hat{\theta} $ when the oscillator is in the state $ |
\psi > $ gives as result $ \theta \in [ 0 , 2 \pi ) $ is:
\begin{equation}
    Pr_{| \psi >}( \theta ) \; := \; | < \theta | \psi > |^{2} \;
    = \; \frac{1}{2 \pi} | \sum_{n=0}^{\infty} c_{n} \exp ( - i n
    \theta ) |^{2}
\end{equation}
Obviously:
\begin{equation} \label{eq:normalization of probability in the bosonic case}
    \int_{0}^{2 \pi} d  \theta \,  Pr_{| \psi >}( \theta ) \; = \; \frac{1}{ 2 \pi
    } \sum_{n=0}^{\infty} \sum_{m=0}^{\infty} c_{n}
    \overline{c_{m}}  \int_{0}^{2 \pi} d \theta \exp [ i ( n - m ) \theta
    ] \; = \;
     \frac{1}{ 2 \pi
    } \sum_{n=0}^{\infty} \sum_{m=0}^{\infty} c_{n}
    \overline{c_{m}} \, 2 \pi \delta_{n,m} \; = \;
    \sum_{n=0}^{\infty} | c_{n}|^{2} \; = \; 1
\end{equation}
where we have used the equation \ref{eq:useful integral}.
\bigskip

\begin{remark}
\end{remark}
The issue concerning the definition of the \emph{phase operator}
is deceptively similar but essentially different from two other
issues:
\begin{enumerate}
    \item the quantization of the dynamical system consisting of a spinless boson of unary mass having as
confiuguration space the circle $ ( S^{1} , \delta := d \theta
\otimes d \theta ) $ and hence having lagrangian $ L : T S^{1}
\mapsto \mathbb{R} $:
\begin{equation}
    L ( \theta , \dot{\theta} ) \; := \; \frac{ |\dot{ \theta} |_{\delta}^{2}
    }{2} \; = \; \frac{ \dot{\theta}^{2} }{2}
\end{equation}
that furnishs the prototypical example of the \emph{topological
superselection rule} with \mbox{\emph{superselection charge} $
\in\ Hom( H_{1}(\text{configuration space}, \mathbb{Z}) , U(1))$}
(discovered independently by Cecile Morette De Witt  and by Larry
Schulman at the end of the sixthes and the beginning of the
seventhes and nowadays commonly founded in the literature: see for
instance the $ 23^{th}$ chapter of \cite{Schulman-81}, the $
7^{th} $ chapter of \cite{Rivers-87} and the $ 8^{th} $ chapter of
\cite{Cartier-De-Witt-Morette-06} as to its implementation, at
different levels of mathematical rigor, in the path-integration's
formulation, as well as the $ 8^{th} $ chapter of
\cite{Balachandran-Marmo-Skagerstam-Stern-91}, the $ 3^{th} $
chapter of \cite{Morandi-92} and the section 6.8 of
\cite{Strocchi-05b} for its formulation in the operatorial
formulation).

Actually, in such a prototypical example, using the \emph{Hurewicz
isomorphism}:
\begin{equation}
    H_{1} (M , \mathbb{Z} ) \; = \; \frac{ \pi_{1} (M) }{ [ \pi_{1} (M) , \pi_{1} (M) ]}
\end{equation}
holding for an arbitrary differentiable manifold M, and where:
\begin{equation}
    [ G , G ] \; := \; \{ a \cdot b \cdot a^{-1} \cdot b^{-1} \; \; a,b \in G   \}
\end{equation}
is the\emph{ commutator subgroup} of an arbitrary group G, if
follows that:
\begin{equation}
  \pi_{1} ( S^{1}) \; = \; \mathbb{Z}
\end{equation}
\begin{equation}
   [ \pi_{1} (S^{1}) , \pi_{1} (S^{1}) ] \; = \; 0
\end{equation}
and hence the involved superselection charge is simply a phase $
\in U(1)$, the distinct superselection sectors  simply
corresponding to different self-adjoint extension of $ -
\frac{1}{2} \frac{d^{2}}{ d \theta^{2} } : C_{0}^{\infty} ( S^{1}
) \mapsto C_{0}^{\infty} ( S^{1} ) $, where $ C_{0}^{\infty} ( M )
$ denotes the set of all the smooth functions with compact support
over an arbitrary differentiable manifold M.
    \item the Bloch theory concerning the lattice $ a
    \mathbb{Z}$ \cite{Sakurai-94} (considered for simplicity in the \emph{tight binding approximation}), i.e. the Quantum Mechanics of a spinless boson of
    unary mass living on the euclidean real line $ ( \mathbb{R} , \delta := dx
\otimes dx ) $ under the influence of a
    field's force with energy potential V(x) periodic of period $ a \in ( 0 , + \infty ) $
    and hence such that:
\begin{equation}
    \tau(a)^{\dag} V(x) \tau(a) \; = \; V( x + a) \; = V(x)
\end{equation}
(where $ \tau ( l ) $ is operator of translation by $ l \in
\mathbb{R}$).

 Obviously the group  $ \{ \tau (x) , x \in a \mathbb{Z} \}$ of the translations by vectors
belonging to the lattice $ a \mathbb{Z}$ is a symmetry of the
system:
\begin{equation}
    [ H , \tau (x) ] \; = \; 0 \; \; \forall  x \in a \mathbb{Z}
\end{equation}
so that H and $ \tau (a) $ may be diagonalized simultaneously.

Denoting with $ | n > $ a  state localized in  the $ n^{th} $ cell
$ [ n a , (n+1) a ] $ and hence such that:
\begin{equation}
    \tau(a) | n > \; = \; | n + 1 >
\end{equation}
the \emph{tight binding approximation} imposes that there exists a
$ \Delta \in ( 0 , + \infty ) $ such that:
\begin{equation}
    < n | H | m > \; = \; - ( \delta_{m,n-1} + \delta_{m,n+1} )
    \Delta \, + \, E_{0} \delta_{n,m}
    \; \; \forall n , m \in \mathbb{Z}
\end{equation}
and hence:
\begin{equation}
    H | n > \; = \; E_{0} | n > - \Delta | n - 1 > - \Delta | n + 1 >
\end{equation}

Introduced the \emph{angle states}:
\begin{equation}
    | \theta > \; := \; \frac{1}{\sqrt{ 2 \pi } } \sum_{n= - \infty}^{ +
    \infty} \exp ( i n \theta ) | n > \; \; \theta \in [ 0 , 2 \pi
    )
\end{equation}
it follows that:
\begin{equation}
  \tau(a) | \theta > \; = \; \frac{1}{\sqrt{ 2 \pi } } \sum_{n= - \infty}^{ +
    \infty} \exp ( i n \theta )  | n + 1 >  \; = \; \sum_{n= - \infty}^{ +
    \infty} \exp [ i ( n- 1)  \theta ]  | n  > \; = \; \exp ( - i
    \theta ) | \theta > \;
    \; \forall \theta \in [ 0 , 2 \pi )
\end{equation}
\begin{multline}
    H  | \theta > \; = \; \frac{1}{\sqrt{ 2 \pi } } \sum_{n= - \infty}^{ +
    \infty} \exp ( i n \theta ) H | n > \; = \\
      \frac{E_{0}}{\sqrt{ 2 \pi } }  \sum_{n= - \infty}^{ +
    \infty} \exp ( i n \theta )  | n > - \frac{\Delta}{\sqrt{ 2 \pi } }  \sum_{n= - \infty}^{ +
    \infty} \exp ( i n \theta )  | n +1 > - \frac{\Delta}{\sqrt{ 2 \pi } }  \sum_{n= - \infty}^{ +
    \infty} \exp ( i n \theta )  | n - 1 >  \; = \\
     \frac{E_{0}}{\sqrt{ 2 \pi } } \sum_{n= - \infty}^{ +
    \infty} \exp ( i n \theta )  | n > - \frac{\Delta}{\sqrt{ 2 \pi } }  \sum_{n= - \infty}^{ +
    \infty} \exp [ i ( n-1) \theta )  | n > - \frac{\Delta}{\sqrt{ 2 \pi } }  \sum_{n= - \infty}^{ +
    \infty} \exp [ i ( n +1)  \theta )  | n  > \; = \\
     ( E_{0} - 2 \Delta \cos ( \theta ) ) | \theta > \;
    \; \forall \theta \in [ 0 , 2 \pi )
\end{multline}
\end{enumerate}

The formal similarities between the issue of defining the phase
operator for a bosonic harmonic oscillator and each of these two
other issues are, anyway,  deceptive since:
\begin{enumerate}
    \item the configuration space of the harmonic
oscillator is the real line having trivial fundamental group. So
no topological superselection rule exists in this case.
    \item the sum in the \emph{angle states} of the periodic
    one-dimensional potential runs from $ - \infty $ to $ + \infty
    $ while the sum in the \emph{bosonic angle states} of the bosonic phase
    operator runs only from 0 to $ + \infty $.

Therefore in the  case of the one dimensional particle in a
periodic energy potential it follows that:
\begin{equation}
  \sum_{n - \infty }^{+ \infty} | n + 1 > < n+1 | \; = \;  \sum_{n - \infty }^{+ \infty} | n  > <  n
  | \; = \; 1
\end{equation}
while in the case of the bosonic phase operator we have seen in
the equation \ref{eq:unitarity failure of the bosonic exponential
phase operator} how the fact that:
\begin{equation}
    \sum_{n=0}^{+ \infty} | n + 1 > < n+1 | \; \neq \; \sum_{n=0}^{+ \infty} | n + 1 > < n+1 |
\end{equation}
\end{enumerate}
is responsible of the fact that the \emph{bosonic exponential
phase operator} is not unitary.

\newpage

\section{Phase operator of the quantum fermionic oscillator}
\label{sec:Phase operator of the quantum fermionic oscillator} Let
us consider a quantum fermionic oscillator having, hence,
hamiltonian:
\begin{equation} \label{eq:hamiltonian of the fermionic oscillator}
    H_{F} \; := \; \omega N_{F}
\end{equation}
where $ a_{F} $ and $ a_{F}^{\dag} $ are respectively the
annihilation and the creation operators:
\begin{equation} \label{eq:1th commutation relation for the fermionic creation and annihilation operators}
    \{  a_{F} ,  a_{F} \} \; = \; \{  a_{F}^{\dag} ,  a_{F}^{\dag} \}
    \; = \; 0
\end{equation}
\begin{equation}  \label{eq:2th commutation relation for the fermionic creation and annihilation operators}
      \{  a_{F} , a_{F}^{\dag} \} \; = \; 1
\end{equation}
and where $ N_{F} $ is the fermionic number operator:
\begin{equation}
   N_{F} \; := \; a_{F}^{\dag} a_{F}
\end{equation}

The equation \ref{eq:1th commutation relation for the fermionic
creation and annihilation operators} and the equation \ref{eq:2th
commutation relation for the fermionic creation and annihilation
operators} imply that:
\begin{equation}\label{eq:commutation relation between fermionic number and annihilation operator}
    [ N_{F} , a_{F} ] \; = \; - a_{F}
\end{equation}
\begin{equation}\label{eq:commutation relation between fermionic number and creation operator}
     [ N_{F} , a_{F}^{\dag} ] \; = \;  a_{F}^{\dag}
\end{equation}
 that imply that:
\begin{equation}
    a_{F} | n > \; = \; \left\{%
\begin{array}{ll}
    0, & \hbox{if $n=0$;} \\
    |0 >, & \hbox{if $ n = 1$.} \\
\end{array}%
\right.
\end{equation}
\begin{equation}
   a_{F}^{\dag} | n > \; = \;  \left\{%
\begin{array}{ll}
    | 1 > , & \hbox{ if $ n = 0$;} \\
    0 , & \hbox{if $ n = 1$.} \\
\end{array}%
\right.
\end{equation}
\begin{equation}
  | n > \; = \; ( a_{F}^{\dag} )^{n} | 0 > \;
  \; \forall n \in \{ 0 , 1 \}
\end{equation}
\begin{equation}
    N_{F} | n > \; = \; n | n > \; \; \forall n \in \{ 0 , 1 \}
\end{equation}
\begin{equation}
    H_{F} | n > \; = \; E_{F} (n) | n >  \; \; \forall n \in \{ 0 , 1 \}
\end{equation}
where:
\begin{equation}  \label{eq:energy levels of the fermionic oscillator}
   E_{F} (n) \; := \; \omega \, n \; \; n \in \{ 0 , 1 \}
\end{equation}

\bigskip

It appears natural, mimicking the approach of the section
\ref{sec:Phase operator of the quantum bosonic oscillator}, to
define the \emph{fermionic angle states} as:
\begin{equation} \label{eq:fermionic angle states}
    | \theta > \; := \; \frac{1}{ \sqrt{2 \pi} }  \sum_{n=0}^{1}
    \exp ( i n \theta ) | n > \; = \;  \frac{1}{ \sqrt{2 \pi} } (
    |0 > + \exp ( i \theta ) | 1 >) \; \; \theta \in [ 0 , 2 \pi )
\end{equation}

\bigskip

\begin{remark}
\end{remark}
Let us remark that:
\begin{equation} \label{eq:scalar product between fermionic angle states}
 < \theta_{1} | \theta_{2} > \; = \; \frac{1}{2 \pi} \{ 1 + \exp [ i ( \theta_{2} -
 \theta_{1} ) ] \}  \; \; \forall \theta_{1} , \theta_{2} \in [ 0 , 2
 \pi)
\end{equation}
and hence in particular:
\begin{equation}
    < \theta | \theta > \; = \; \frac{1}{\pi} \; \; \forall \theta  \in [ 0 , 2
 \pi)
\end{equation}
Though not orthonormal, the \emph{fermionic angle states} are
complete:
\begin{equation} \label{eq:completeness of the fermionic angle states}
    \int_{0}^{2 \pi} d \theta | \theta > < \theta | \; = \frac{1}{2 \pi} \; \int_{0}^{2
    \pi} d \theta ( | 0 > < 0 | + \exp ( - i \theta ) | 0 > < 1 | +  \exp (  i \theta ) | 1 > < 0 |
    + | 1 > < 1 | ) \; = \; 1
\end{equation}

\bigskip

Always mimicking the approach of the section \ref{sec:Phase
operator of the quantum bosonic oscillator}, it would then appear
natural to define the \emph{fermionic  exponential phase operator}
as the operator $  | 0 > < 1 | $.

Anyway:
\begin{equation} \label{eq:failure of the naife fermionic exponential operator}
   | 0 > < 1 | \theta > \; = \; \frac{1}{\sqrt{2 \pi} } ( | 0 > < 1 | 0 > + \exp ( i \theta ) | 0 > < 1 | 1
   > )  \; = \;  \frac{ \exp ( i \theta )  }{\sqrt{2 \pi}}  | 0 > \; \neq \; \exp ( i \theta ) | \theta >
\end{equation}
and hence $ | \theta > $ is not an eigenstate of $  | 0 > < 1 | $.

Let us then proceed in  a different way expressing the
\emph{exponential phase operator} in the more general way:
\begin{equation}
  \exp ( i \hat{\theta} ) \; := \; c_{00} |0 > < 0 | + c_{01} |0 > < 1
  | + c_{10} |1 > < 0 | + c_{11} |1 > < 1 |
\end{equation}
and imposing the condition:
\begin{equation}
     \exp ( i \hat{\theta} )  | \theta >  \; = \; \exp ( i \theta
     ) | \theta > \; \; \forall \theta \in [ 0 , 2 \pi )
\end{equation}
and hence that:
\begin{equation}
    \exp ( i \hat{\theta} )  | \theta >  \; = \; \frac{1}{\sqrt{2 \pi}} ( \exp ( i \theta
     ) | 0 > + \exp ( i 2 \theta
     ) | 1 > ) \; \; \forall \theta \in [ 0 , 2 \pi )
\end{equation}
Since:
\begin{equation}
  \exp ( i \hat{\theta} )  | \theta >  \; = \; \frac{1}{\sqrt{2 \pi}} [ ( c_{00} + \exp ( i \theta
     ) )  | 0 > + ( c_{10} + c_{11}  \exp ( i \theta )
     ) | 1 > ]
\end{equation}
it follows that:
\begin{equation} \label{eq:1th condidition on the coefficients}
    c_{00} + c_{01} \exp ( i \theta ) \; = \; \exp ( i \theta )
\end{equation}
\begin{equation} \label{eq:2th condidition on the coefficients}
    c_{10} + c_{11}  \exp ( i \theta ) \; = \; \exp ( i 2 \theta )
\end{equation}
The imposition of the unitarity of the \emph{fermionic exponential
phase operator} leads to the constraints:
\begin{equation}\label{eq:1th unitarity constraint}
    |c_{0,0}|^{2} + | c_{1,0} |^{2} \; = \; 1
\end{equation}
\begin{equation}\label{eq:2th unitarity constraint}
    | c_{0,1}|^{2} + | c_{1,1} |^{2} \; = \; 1
\end{equation}
\begin{equation}\label{eq:3th unitarity constraint}
    \overline{c_{0,0}} c_{0,1}  + \overline{c_{1,0}} c_{1,1} \; = \; 0
\end{equation}
\begin{equation}\label{eq:4th unitarity constraint}
    c_{0,0} \overline{c_{0,1}}  + c_{1,0} \overline{c_{1,1}} \; = \; 0
\end{equation}
\begin{equation}\label{eq:5th unitarity constraint}
    | c_{0,0} |^{2} + | c_{0,1} |^{2} \; = \; 1
\end{equation}
\begin{equation}\label{eq:6th unitarity constraint}
    | c_{1,0}| ^{2} + | c_{1,1} |^{2} \; = \; 1
\end{equation}
\begin{equation}\label{eq:7th unitarity constraint}
    c_{0,0} \overline{c_{1,0}}  + c_{0,1} \overline{c_{1,1}} \; = \; 0
\end{equation}
\begin{equation}\label{eq:8th unitarity constraint}
   \overline{ c_{0,0}} c_{1,0}  + c_{1,1} \overline{c_{0,1}} \; = \; 0
\end{equation}

We will now show that that there don't exist four complex numbers
$ c_{00} , c_{01} , c_{10} , c_{11} $ satisfying simultaneously
the equation \ref{eq:1th condidition on the coefficients}, the
equation \ref{eq:2th condidition on the coefficients}, the
equation \ref{eq:1th unitarity constraint}, the equation
\ref{eq:2th unitarity constraint}, the equation \ref{eq:3th
unitarity constraint}, the equation \ref{eq:4th unitarity
constraint}, the equation \ref{eq:5th unitarity constraint}, the
equation \ref{eq:6th unitarity constraint}, the equation
\ref{eq:7th unitarity constraint} and  the equation \ref{eq:8th
unitarity constraint}.

Given a normalized state:
\begin{equation} \label{eq:generic fermionic state}
    | \psi > \; := \; d_{0} |0 > + d_{1} | 1 >
\end{equation}
\begin{equation} \label{eq:fermionic normalization condition}
    < \psi | \psi > \; = \; | d_{0} |^{2} +  | d_{1} |^{2} \; = \;
    1
\end{equation}
it follows that:
\begin{multline}
    < \psi | ( \exp ( i \hat{\theta} ))^{\dag} \exp ( i \hat{\theta}
    ) | \psi > \; = \;   \int_{0}^{2 \pi} d \theta \int_{0}^{2 \pi} d \theta'  < \psi |
    \theta > < \theta | ( \exp ( i \hat{\theta} ))^{\dag}  \exp ( i \hat{\theta}
    )  |\theta' > < \theta' | \psi > \; = \\
     \int_{0}^{2 \pi} d \theta \int_{0}^{2 \pi} d
    \theta' \exp [  i ( \theta' - \theta ) ]   < \psi | \theta > <
    \theta | \theta' > <  \theta' | \psi >
\end{multline}
where we have used the completeness of the fermionic angle states
stated by the equation \ref{eq:completeness of the fermionic angle
states}.

Since:
\begin{equation}
    < \psi | \theta > \; = \; ( < 0 | \overline{d_{0}} + < 1 |
    \overline{d_{1}} ) \frac{1}{ \sqrt{2 \pi} } ( | 0 > + \exp ( i
    \theta ) | 1 > )\; = \; \frac{1}{ \sqrt{2 \pi} } (
    \overline{d_{0}} +  \exp ( i \theta ) \overline{d_{1}})
\end{equation}
\begin{equation}
    < \theta' | \psi > \; = \; \frac{1}{ \sqrt{2 \pi} } ( d_{0} +
    \exp( - i \theta') d_{1} )
\end{equation}
it follows that:
\begin{multline}
   < \psi | ( \exp ( i \hat{\theta} ))^{\dag} \exp ( i \hat{\theta}
    ) | \psi > \; = \\
     \frac{1}{ ( 2 \pi )^{2} } \int_{0}^{ 2 \pi} d
    \theta \int_{0}^{ 2 \pi} d
    \theta' \{ \exp [ i ( \theta' - \theta ) ] +  \exp [2 i ( \theta' - \theta )
    ] \} \{ | d_{0}|^{2} + \overline{ d_{0}} d_{1} \exp ( - i
    \theta' ) + d_{0} \overline{d_{1}} \exp ( i \theta ) + |
    d_{1}|^{2} \exp [ i ( \theta - \theta' )] \} \; = \\
 \frac{ | d_{1} |^{2}  }{ ( 2 \pi )^{2} } \; \neq \; 1
\end{multline}
(where we have used the equation \ref{eq:scalar product between
fermionic angle states}) and hence  the \emph{fermionic
exponential phase operator} $ \exp ( i \hat{\theta} ) $ is not
unitary and the \emph{fermionic phase operator} $ \hat{\theta} $
is not self-adjoint.

As we have seen this is equivalent to the fact that there don't
exist four complex numbers $ c_{00} , c_{01} , c_{10} , c_{11} $
satisfying simultaneously the equation \ref{eq:1th condidition on
the coefficients}, the equation \ref{eq:2th condidition on the
coefficients}, the equation \ref{eq:1th unitarity constraint}, the
equation \ref{eq:2th unitarity constraint}, the equation
\ref{eq:3th unitarity constraint}, the equation \ref{eq:4th
unitarity constraint}, the equation \ref{eq:5th unitarity
constraint}, the equation \ref{eq:6th unitarity constraint}, the
equation \ref{eq:7th unitarity constraint} and  the equation
\ref{eq:8th unitarity constraint}.

Imposing only the  equation \ref{eq:1th condidition on the
coefficients} and the equation \ref{eq:2th condidition on the
coefficients} one obtains many possible solutions among which
there is the choice:
\begin{equation}
    c_{00} \; := \; 0
\end{equation}
\begin{equation}
    c_{01} \; := \; 1
\end{equation}
\begin{equation}
    c_{10} \; := \; \exp ( i \theta )
\end{equation}
\begin{equation}
   c_{11} \; := \; 1
\end{equation}
determining the \emph{fermionic exponential phase operator}:
\begin{equation} \label{eq:fermionic exponential phase operator}
  \exp ( i \hat{\theta} ) \; = \; | 0 > < 1 | + \exp ( i \theta )  | 1 > < 0
  | +  | 1 > < 1 |
\end{equation}

\bigskip

Given the generic normalized state given by the equation
\ref{eq:generic fermionic state} and the equation
\ref{eq:fermionic normalization condition}, the probability that a
measurement of the \emph{fermionic phase operator} $ \hat{\theta}
$ when the oscillator is in the state $ | \psi > $ gives as result
$ \theta \in [ 0 , 2 \pi ) $ is:
\begin{equation}
    Pr_{| \psi >}( \theta ) \; := \; | < \theta | \psi > |^{2} \;
    = \; \frac{1}{2 \pi} ( 1 + d_{0} \overline{d_{1}} \exp ( i \theta ) +
    \overline{d_{0}} d_{1} \exp ( - i \theta ) )
\end{equation}
Obviously:
\begin{equation} \label{eq:normalization of probability in the fermionic case}
    \int_{0}^{2 \pi} d \theta \, Pr_{| \psi >}( \theta ) \; = \; \frac{1}{2
    \pi} + ( 2 \pi + d_{0} \overline{d_{1}} \int_{0}^{2 \pi} d \theta \exp ( i \theta
    ) + \overline{d_{0}} d_{1} \int_{0}^{2 \pi} d \theta \exp ( i \theta
    ) ) \; = \; 1
\end{equation}

\newpage
\section{Phase properties of the quantum supersymmetric
oscillator}
Let us now consider the quantum supersymmetric
oscillator (see for instance the $ 6^{th} $ chapter
"Supersymmetry" of \cite{Das-93}) having hamiltonian:
\begin{equation} \label{eq:hamiltonian of the supersymmetric oscillator}
    H \; := \; H_{B} + H_{F}
\end{equation}
where $ H_{B} $ and $ H_{F} $ are the hamiltonians
of, respectively, the \emph{quantum bosonic oscillator} and the
\emph{quantum fermionic oscillator} given, respectively, by the
equation \ref{eq:hamiltonian of the bosonic oscillator} and the
equation \ref{eq:hamiltonian of the fermionic oscillator}, and
where:
\begin{equation}
    [ a_{B} , a_{F} ] \; = \; [ a_{B}^{\dagger} , a_{F} ] \; = \; [ a_{B} , a_{F}^{\dagger}
    ] \; = \; [ a_{B}^{\dagger} , a_{F}^{\dagger} ] \; = \; 0
\end{equation}
Clearly:
\begin{equation}
    H | n_{B} , n_{F} > \; = E( n_{B} , n_{F} ) | n_{B} , n_{F} >
    \; \; \forall n_{B} \in \mathbb{N} , \forall n_{F} \in \{ 0 ,
    1 \}
\end{equation}
where:
\begin{equation}
  E( n_{B} , n_{F} ) \; := E_{B} ( n_{B} ) + E_{F} ( n_{F} ) \; \;
  n_{B} \in \mathbb{N} ,  n_{F} \in \{ 0 ,
    1 \}
\end{equation}
\begin{equation}
    | n_{B} , n_{F} > \; = \; \frac{ ( a_{B} ^{\dag} )^{n_{B}}  }{ \sqrt{ n_{B} ! }
    } ( a_{F} ^{\dag} )^{n_{F}} | 0 > \; \; \forall n_{B} \in \mathbb{N} , \forall n_{F} \in \{ 0 ,
    1 \}
\end{equation}
with $ E_{B} ( n_{B} ) $ and  $ E_{F} ( n_{F} ) $ defined,
respectively, by the equation \ref{eq:energy levels of the bosonic
oscillator} and the equation \ref{eq:energy levels of the
fermionic oscillator}.

Let us now introduce the operators:
\begin{equation} \label{eq:1th susy charge}
    Q \; := \; a_{B}^{\dag} a_{F}
\end{equation}
\begin{equation} \label{eq:2th susy charge}
    \bar{Q} \; := Q^{\dag} \; = \; a_{F}^{\dag} a_{B}
\end{equation}
Since:
\begin{equation} \label{eq:the supercharges commute with the hamiltonian}
    [ Q , H ] \; = \; [ \bar{Q} , H ] \; = \; 0
\end{equation}
Q and $ \bar{Q} $ are symmetries of the quantum supersymmetric
oscillator and:
\begin{equation}
    H ( Q | n_{B} , n_{F} > ) \; =  \; Q H | n_{B} , n_{F} > \; = \; E ( n_{B} , n_{F} ) ( Q | n_{B} , n_{F} >
    ) \; \; \forall n_{B} \in \mathbb{N} , \forall n_{F} \in \{ 0 ,
    1 \}
\end{equation}
\begin{equation}
    H ( \bar{Q} | n_{B} , n_{F} > ) \; = \; \bar{Q} H  | n_{B} , n_{F} >  \; = \;  E ( n_{B} , n_{F} ) ( \bar{Q} | n_{B} , n_{F} >
    ) \; \; \forall n_{B} \in \mathbb{N} , \forall n_{F} \in \{ 0 ,
    1 \}
\end{equation}
It may be, furthermore, easily verified that:
\begin{equation}
    [ N_{B} ,  \bar{Q} ] \; = \; - \bar{Q}
\end{equation}
\begin{equation}
    [  N_{B} , Q ] \; = \;  Q
\end{equation}
\begin{equation}
    [ N_{F} , Q ] \; = \; - Q
\end{equation}
\begin{equation}
    [ N_{F} , \bar{Q} ] \; = \;  \bar{Q}
\end{equation}
from which it follows that:
\begin{equation} \label{eq:action of the 1th susy charge}
    Q | n_{B} , n_{F} > \; = \; \left\{%
\begin{array}{ll}
    \sqrt{ n_{B} +1 } | n_{B} + 1 , n_{F} -1 >  , & \hbox{if $ n_{F}=1 $;} \\
    0, & \hbox{if $ n_{F} = 0$.} \\
\end{array}%
\right.
\end{equation}
\begin{equation} \label{eq:action of the 2th susy charge}
    \bar{Q} | n_{B} , n_{F} > \; = \; \left\{%
\begin{array}{ll}
   \sqrt{n_{B}} | n_{B} - 1 , n_{F} +1 >  , & \hbox{if $ n_{B} \in \mathbb{N}_{+} $ and $ n_{F} = 0 $;} \\
    0, & \hbox{if $ n_{B} = 0$ or $ n_{F} = 1 $.} \\
\end{array}%
\right.
\end{equation}

Since, informally speaking, one can say that Q transforms a
"fermionic quantum" into a "bosonic quantum" while $ \bar{Q}$
transforms a "bosonic quantum" into a "fermionic quantum", Q and $
\bar{Q} $ are called \emph{supersymmetric charges}.

\bigskip

The approach followed in the section \ref{sec:Phase operator of
the quantum bosonic oscillator}
 and in the section \ref{sec:Phase operator of the quantum fermionic oscillator} leads  naturally to define the \emph{supersymmetric angle states} as:
\begin{equation} \label{eq:supersymmetric angle states}
    | \theta > \; := \; \frac{1}{\sqrt{2}} (| \theta >_{B} \otimes  | \theta >_{F}) \; = \; \frac{1}{(2)^{\frac{3}{2}} \pi} \sum_{n_{B}=0}^{\infty} \sum_{n_{F}=0}^{1}
    \exp [ i ( n_{B} + n_{F} ) \theta ] | n_{B} , n_{F} > \; \;
    \theta \in [ 0 , 2 \pi )
\end{equation}
where $  | \theta >_{B} $ and $  | \theta >_{F} $ are,
respectively, the \emph{bosonic angle state}  and the \emph{
fermionic angle state}  defined, respectively, by the equation
\ref{eq:bosonic angle states} and by the equation
\ref{eq:fermionic angle states}.

\bigskip

\begin{remark}
\end{remark}
Let us remark that:
\begin{equation}
    < \theta_{1} | \theta_{2} > \; = \; \frac{1}{8 \pi^{2}}  \sum_{n_{B}=0}^{\infty}
    \sum_{n_{F}=0}^{1} \exp \{  i [ ( n_{B} + n_{F}) ( \theta_{2}
    - \theta_{1} ) ] \} \; \neq \; \delta ( \theta_{1} -
    \theta_{2})
\end{equation}
though obviously:
\begin{equation}
     < \theta | \theta > \; = \; + \infty
\end{equation}
Though not orthonormal, the \emph{supersymmetric angle states} are
complete:
\begin{equation}
    \int_{0}^{2 \pi} d \theta | \theta > < \theta | \; =  \int_{0}^{2 \pi} d \theta ( | \theta>_{B} < \theta
    |_{B}  \otimes 1_{F} ) +   \int_{0}^{2 \pi} d \theta ( 1_{B} \otimes | \theta >_{F} < \theta
    |_{F} ) \; = \; 1
\end{equation}
where we have used the completeness condition of, respectively,
the \emph{bosonic angle states} and the\emph{ fermionic angle
states} given, respectively, by the equation \ref{eq:completeness
of the bosonic angle states} and by the equation
\ref{eq:completeness of the fermionic angle states}.

\bigskip

It would appear natural to define the \emph{supersymmetric
exponential phase operator} as \mbox{ $ \sum_{n_{B}=0}^{\infty} |
n_{B} , 0 >  < n_{B} + 1 , 1 |$}.

Anyway the same considerations concerning the \emph{fermionic
exponential phase operator} and condensed in the equation
\ref{eq:failure of the naife fermionic exponential operator} lead
us to observe that:
\begin{equation}
  \sum_{n_{B}=0}^{\infty} | n_{B} ,
0 >  < n_{B} + 1 ,  1 | \theta > \; = \;
\frac{1}{(2)^{\frac{3}{2}} \pi} \sum_{n_{B} =0}^{\infty} \exp [ i
( n_{B} + 2 ) \theta ] | n_{B} , 0 > \; \neq \; \exp ( i \theta )
| \theta >
\end{equation}

Since in the last section we have, indeed, seen that the correct
\emph{fermionic exponential phase operator} is given by the
equation \ref{eq:fermionic exponential phase operator} it follows
that the \emph{supersymmetric exponential phase operator} is:
\begin{multline} \label{eq:supersymmetric exponential phase operator}
    \exp ( i \hat{\theta} ) \; := \; \exp ( i \hat{\theta}_{B} ) \otimes \exp ( i \hat{\theta}_{F} ) \; = \;  ( \sum_{n=0}^{\infty} | n > < n+1 |
    ) \otimes ( | 0 > < 1 | + \exp ( i \theta ) | 1 > < 0 | + | 1
    > < 1 | ) \; = \\
     \sum_{n=0}^{ \infty } ( | n,0 > < n+1 ,1 |+
    \exp( i \theta ) | n,1 > < n+1 ,0 | + | n , 1 >< n+1 , 1 |)
\end{multline}
since it obeys the equation:
\begin{equation}
  \exp ( i \hat{\theta} )  | \theta >  \; = \; \frac{1}{\sqrt{2}} [ \exp ( i \hat{\theta}_{B} ) \otimes \exp ( i \hat{\theta}_{F} ) ]  | \theta >_{B} \otimes | \theta >_{F}  \; = \; \exp ( i \theta ) | \theta >     \; \; \forall \theta \in [ 0 , 2 \pi )
\end{equation}

\bigskip

\begin{remark}
\end{remark}
Let us remark that, by construction, the \emph{supersymmetric
exponential phase operator} $ \exp ( i \hat{\theta} )  $  is not
unitary and hence the \emph{supersymmetric phase operator} $
\hat{\theta} $ is not self-adjoint.

\bigskip

Given a normalized product state:
\begin{equation}
    | \psi > \; := \; | \psi >_{B} \otimes | \psi >_{F}
\end{equation}
\begin{equation}
  | \psi >_{B} \; := \;  \sum_{n_{B}=0}^{\infty} c_{n_{B}} | n_{B} >
\end{equation}
\begin{equation}
  \sum_{n_{B}=0}^{\infty} | c_{n_{B}} |^{2} \; = \; 1
\end{equation}
\begin{equation}
    | \psi >_{F} \; := \;  \sum_{n_{F}=0}^{1} c_{n_{F}} | n_{F} >
\end{equation}
\begin{equation}
    \sum_{n_{F}=0}^{1} | c_{n_{F}} |^{2} \; = \; 1
\end{equation}
\begin{equation}
  < \psi | \psi > \; = \;   < \psi |_{B} | \psi >_{B} \, < \psi |_{F} | \psi >_{F} \; = \; 1
\end{equation}
\begin{equation}
   < \theta | \psi > \; =   < \theta |_{B} | \psi >_{B} \, < \theta |_{F} | \psi >_{F}
\end{equation}
let us introduce the following two events:
\begin{itemize}
    \item $ EV_{B}( | \psi > , \theta ) := $ "a measurement of the
\emph{bosonic phase operator} $ \hat{\theta}_{B} $, when the the
supersymmetric oscillator is in the state $ | \psi
> $, gives as result $ \theta \in [ 0 , 2 \pi ) $"
    \item  $ EV_{F}( | \psi > , \theta ) := $ "a measurement of the
\emph{fermionic phase operator} $ \hat{\theta}_{F} $, when the
supersymmetric oscillator is in the state $ | \psi
> $, gives as result $ \theta \in [ 0 , 2 \pi ) $"
\end{itemize}
The fact that $ | \psi > $ is a \emph{product state} implies that
$  EV_{B}( | \psi > , \theta ) $ and $ EV_{F}( | \psi > , \theta )
$ are \emph{independent events} and hence:
\begin{multline}
    Pr[  EV_{B}( | \psi > , \theta ) \wedge  EV_{F}( | \psi > , \theta
    ) ] \; = \;  Pr[  EV_{B}( | \psi > , \theta ) ] \cdot Pr[  EV_{F}( | \psi > , \theta
    ) ] \; = \\
     Pr_{| \psi >_{B} } ( \theta ) \cdot   Pr_{| \psi >_{F}} ( \theta ) \; =
    \; |  < \theta |_{B} | \psi >_{B} |^{2} \cdot |  < \theta |_{F} | \psi >_{F}
    |^{2} \; = \; |  < \theta |  \psi > |^{2} \; \; \forall \theta
    \in [ 0 , 2 \pi)
\end{multline}
Obviously:
\begin{multline}
    \int_{0}^{2 \pi} d \theta_{1}  \int_{0}^{2 \pi} d \theta_{2}  Pr[  EV_{B}( | \psi > , \theta_{1} ) \wedge  EV_{F}( | \psi > , \theta_{2}
    ) ] \; = \; \int_{0}^{2 \pi} d \theta_{1}  \int_{0}^{2 \pi} d
    \theta_{2} Pr[  EV_{B}( | \psi > , \theta _{1}) ] \cdot Pr[  EV_{F}( | \psi > , \theta_{2}
    ) ] \; = \\
 ( \int_{0}^{2 \pi} d \theta_{1} Pr[  EV_{B}( | \psi > , \theta _{1}) ]
 ) \cdot ( \int_{0}^{2 \pi} d \theta_{2} Pr[  EV_{F}( | \psi > , \theta _{2}) ]
 ) \; = \; [ \int_{0}^{2 \pi} d \theta_{1} Pr_{|\psi>_{B}} (
 \theta_{1}) ] \cdot [ \int_{0}^{2 \pi} d \theta_{2} Pr_{|\psi>_{F}} (
 \theta_{2}) ] \; = \; 1
 \end{multline}
 where we have used the equation \ref{eq:normalization of probability in the bosonic case}
 and the equation \ref{eq:normalization of probability in the fermionic
 case}.

\smallskip

When the state $ | \psi > $ is \emph{entangled}, $  EV_{B}( | \psi
> , \theta ) $ and $ EV_{F}( | \psi > , \theta ) $ are not
\emph{independent events} so that:
\begin{equation}
     Pr[  EV_{B}( | \psi > , \theta ) \wedge  EV_{F}( | \psi > , \theta
    ) ] \; \neq \;  Pr[  EV_{B}( | \psi > , \theta ) ] \cdot Pr[  EV_{F}( | \psi > , \theta
    ) ]
\end{equation}
and the situation is more complex.

\bigskip

Let us now finish to consider separately  measurements of the
\emph{bosonic phase operator} and of the \emph{fermionic phase
operator} and let us take into account directly measurement of the
\emph{supersymmetric phase operator}.

Given a normalized state:
\begin{equation}
    | \psi > \; := \; \sum_{n_{B}=0}^{\infty} \sum_{n_{F}=0}^{1}
    c_{n_{B},n_{F}} | n_{B} , n_{F} >
\end{equation}
\begin{equation} \label{eq:supersymmetric normalization condition}
  \sum_{n_{B}=0}^{\infty} \sum_{n_{F}=0}^{1} | c_{n_{B},n_{F}}
  |^{2} \; = \; 1
\end{equation}
satisfying the following mysterious constraint:
\begin{equation} \label{eq:mysterious constraint}
    \sum_{n_{B}=0}^{\infty} \Re [ c_{n_{B},1}
    \overline{c_{n_{B}+1,0}} ] \; = \; 0
\end{equation}
the interpretation of $ | < \theta | \psi
> |^{2} $ as the probability that a measurement of the \emph{supersymmetric phase
operator}, when the supersymmetric oscillator is the state $ |
\psi >$, gives as result $ \theta \in [ 0 , 2 \pi ) $:
\begin{equation}
    Pr_{| \psi >}( \theta ) \; := \; | < \theta | \psi > |^{2} \;
    = \sum_{n_{B}=0}^{\infty} \sum_{n_{F}=0}^{1} \sum_{n'_{B}=0}^{\infty}
    \sum_{n'_{F}=0}^{1} \exp [ i \theta ( n_{B} + n_{F} - n'_{B} -
    n'_{F} ) ] \overline{c_{n_{B},n_{F}}} c_{n'_{B},n'_{F}}
\end{equation}
is consistent since:
\begin{multline}
  \int_{0}^{ 2 \pi} d \theta Pr_{| \psi >} ( \theta ) \; = \;
  \sum_{n_{B}=0}^{\infty} \sum_{n_{B}'=0}^{\infty}
  \int_{0}^{\infty} d \theta \exp [ i ( n_{B} - n_{B}' ) \theta ]
  (\overline{c_{n_{B},0}}  c_{n_{B}',0} + \overline{c_{n_{B},1}}
  c_{n_{B}',1}) + \\
 \sum_{n_{B}=0}^{\infty} \sum_{n_{B}'=0}^{\infty}
  \int_{0}^{\infty}  d \theta \exp [ i ( n_{B} - n_{B}' +1 ) \theta
  ] c_{n_{B},1} \overline{c_{n_{B}',0}} + \sum_{n_{B}=0}^{\infty} \sum_{n_{B}'=0}^{\infty}
  \int_{0}^{\infty}  d \theta \exp [ i ( n_{B} - n_{B}' - 1 ) \theta
  ] c_{n_{B},0} \overline{c_{n_{B}',1}} \; = \\
   \sum_{n_{B}=0}^{\infty}  ( | c_{n_{B},0} |^{2} + | c_{n_{B},1}
   |^{2} ) + \sum_{n_{B}=0}^{\infty} (c_{n_{B},1} \overline{c_{n_{B}+1,0}}
    + c_{n_{B}+1,0} \overline{c_{n_{B},1}}) \; = \; 1
\end{multline}
where we have used the equation \ref{eq:useful integral}, the
equation \ref{eq:supersymmetric normalization condition} and the
mysterious constraint of the equation \ref{eq:mysterious
constraint}.

Contrary, if the mysterious constraint of the equation
\ref{eq:mysterious constraint} is not satisfied, such a
probabilistic interpretation is not consistent.

Let us introduce the set of the states of $ \mathcal{H} :=
\mathcal{H}_{B} \otimes \mathcal{H}_{F} $  satisfying such a
constraint:
\begin{equation}
    \mathcal{H}_{constraint} \; := \; \{ | \psi > =
    \sum_{n_{B}=0}^{\infty} \sum_{n_{F}=0}^{1} c_{n_{B},n_{F}} |
    n_{B}, n_{F} > \in \mathcal{H} \;
    : \\
  \sum_{n_{B}=0}^{\infty} \Re [ c_{n_{B},1} \overline{c_{n_{B}+1,0}}
] \; = \; 0 \}
\end{equation}
It may be easily verified that:
\begin{enumerate}
    \item $ \mathcal{H}_{constraint} $ is not a linear subspace of
     $ \mathcal{H} $.
    \item its complement $ \mathcal{H}
    - \mathcal{H}_{constraint} $ contains both \emph{product states} and
    \emph{entangled states}, i.e.:
\begin{equation}
   \mathcal{H}_{constraint} \cap \mathcal{H}_{product} \; \neq \;  \emptyset
\end{equation}
\begin{equation}
     \mathcal{H}_{constraint} \cap \mathcal{H}_{entangled} \; \neq \;  \emptyset
\end{equation}
where obviously:
\begin{equation}
  \mathcal{H}_{product} \; := \;\{ | \psi >_{B} \otimes  | \psi  >_{F}  \; \;  | \psi >_{B} \in \mathcal{H}_{B} ,  | \psi >_{F} \in
  \mathcal{H}_{F} \}
\end{equation}
\begin{equation}
  \mathcal{H}_{entangled} \; := \; \mathcal{H} - \mathcal{H}_{product}
\end{equation}
\end{enumerate}
\newpage
\part{Theory at strictly positive temperature.}
\section{A brief review of Umezawa's thermofield dynamics}

Among the different existing approaches available to study quantum
field theories at strictly positive temperature
\cite{Le-Bellac-96}, \cite{Kapusta-Gale-06}, H. Umezawa's
approach, usually called \emph{thermofield dynamics}, is
particularly adapted to the discussion of symmetry breaking
issues, as we will briefly recall following closely the $ 3^{th} $
chapter "Thermofield Dynamics" of \cite{Das-97} and
\cite{Umezawa-95}.

Given a quantum system having an hamiltonian H (being of course a
self-adjoint operator over a suitable Hilbert space $ \mathcal{H}
$) with discrete spectrum:
\begin{equation}
    H | n > \; = \; E_{n} | n >
\end{equation}
\begin{equation}
    < n | m > \; = \; \delta_{n,m}
\end{equation}
being in thermodynamical equilibrium with a thermal bath at
temperature $ T > 0 $ , let us define a \emph{thermal vacuum at
inverse temperature $ \beta := \frac{1}{T}$} as a state $ | 0 ;
\beta > $ such that the expectation value $ < 0 ; \beta | A |0 ;
\beta > $ of an arbitrary observable A is equal to the statistical
average of A over the \emph{canonical ensemble}, i.e.:
\begin{equation} \label{eq:basic property of the thermal vacuum}
    < 0 ; \beta | A | 0 ; \beta > \; =  < A >_{\beta} \; = \; \frac{ Tr \exp ( - \beta H ) A }
    {Z(\beta)} \; = \; \frac{ \sum_{n} \exp ( - \beta E_{n}) < n | A | n >  }{ Z(\beta) }
\end{equation}
where:
\begin{equation}
    Z( \beta ) \; := \; Tr \exp ( - \beta H )
\end{equation}
is the \emph{canonical partition function}.

Using the completeness condition for the eigenvectors of the
hamiltonian:
\begin{equation}
  \sum_{n} | n > < n | \; = \; 1
\end{equation}
we obtain that:
\begin{equation}
    | 0 ; \beta > \; = \;  \sum_{n} | n > < n | 0 ; \beta >
\end{equation}
\begin{equation}
    < 0 ; \beta | \; = \;  \sum_{m} < 0 ; \beta | m > < m |
\end{equation}
 and hence we can write the expectation value of the
observable A over the \emph{thermal vacuum} as:
\begin{equation}
   < 0 ; \beta | A | 0 ; \beta > \; = \; \sum_{n} \sum_{m}  < n | 0 ; \beta
   >  < 0 ; \beta | m > < m | A | n >
\end{equation}
that since:
\begin{equation}
     < 0 ; \beta | m > \; = \; \overline{ < m | 0 ; \beta >}
\end{equation}
becomes:
\begin{equation} \label{eq:expectation value over the thermal vacuum in the non-doubled Hilbert space}
     < 0 ; \beta | A | 0 ; \beta > \; = \; \sum_{n} \sum_{m} < n | 0 ; \beta
     >  \overline{ < m | 0 ; \beta >} < m | A | n >
\end{equation}
Comparing the equation \ref{eq:basic property of the thermal
vacuum} with the equation \ref{eq:expectation value over the
thermal vacuum in the non-doubled Hilbert space} we see that a
\emph{thermal vacuum} $ | 0 ; \beta > \in \mathcal{H} $ should
satisfy the impossible condition:
\begin{equation}
   < n | 0 ; \beta >  \overline{ < m | 0 ; \beta >} \; = \; \frac{ \exp( - \beta E_{n}) \delta_{n,m}}{Z(\beta) }
\end{equation}

Hence a \emph{thermal vacuum} $ | 0 ; \beta > \in \mathcal{H} $
doesn't exist.

It follows that, if we insist on looking for a \emph{thermal
vacuum}, we have to search it in a suitably enlarged Hilbert
space.

The simplest choice is the doubled Hilbert space $ \mathcal{H}
\otimes \tilde{\mathcal{H}} $ where $ \tilde{\mathcal{H}} \; :=
\mathcal{H} $ is a copy of $  \mathcal{H} $.

Let us denote with $  | \tilde{n} > $  the identical copy of the
vector $ | n> $ but belonging to the copy Hilbert space $
\tilde{\mathcal{H}} $.

Obviously:
\begin{equation}
  \sum_{n} | n > < n | \; = \; 1_{\mathcal{H}}
\end{equation}
\begin{equation}
   \sum_{n} | \tilde{n} > < \tilde{ n }| \; = \; 1_{\tilde{\mathcal{H}}}
\end{equation}
\begin{equation}
  \sum_{n}  \sum_{m} | n , \tilde{m} > < n , \tilde{m} | \; = \; 1_{\mathcal{H} \otimes \tilde{\mathcal{H}}}
\end{equation}
\begin{equation} \label{eq:orthormality condition in the doubled Hilbert space}
    < n , \tilde{m} | n' , \tilde{m}' > \; = \; \delta_{n, n'} \delta_{m,m'}
\end{equation}

 Hence we can express the putative \emph{thermal vacuum} as:
\begin{equation}
    | 0 ; \beta > \; = \sum_{n} \sum_{\tilde{m}}  | n , \tilde{m} > < n , \tilde{m}
    | 0 ; \beta >
\end{equation}
Let us now observe that since the copy orthonormal basis $ \{ |
\tilde{n} > \} $ of the copy Hilbert space $ \tilde{\mathcal{H}} $
is identical to the basis $ \{ | n > \} $ of $ \mathcal{H} $, it
follows that:
\begin{equation}
    <  n , \tilde{m} | 0 ; \beta > \; = \; \delta_{n, \tilde{m} }
    < n ,  \tilde{m} | 0 ; \beta >
\end{equation}
and hence:
\begin{equation}
    | 0 ; \beta > \; = \; \sum_{n} | n , \tilde{n} >  < n , \tilde{n} | 0 ; \beta >
\end{equation}

\bigskip

\begin{remark}
\end{remark}
Let us remark that given an observable of our system, i.e. a
self-adjoint operator A over $ \mathcal{H} $:
\begin{equation} \label{eq:1th decoupling condition}
    < n , \tilde{m} | A | n' ,  \tilde{m}' > \; = \; < n | A | n' > <
    \tilde{m} | \tilde{m}' > \; = \; < n | A | n' > \delta_{\tilde{m}, \tilde{m}'}
\end{equation}
Considering instead the corresponding operator $ \tilde{A} $ over
the copy Hilbert space $ \tilde{\mathcal{H}} $:
\begin{equation}\label{eq:2th decoupling condition}
  <  n , \tilde{m} | \tilde{A} | n' ,  \tilde{m}' > \; = \; < n | n'
  > < \tilde{m} | \tilde{A} | \tilde{m}' > \; = \delta_{n,n'} < \tilde{m} | \tilde{A} | \tilde{m}' >
\end{equation}

\bigskip

Given an observable A of our system we have then that:
\begin{multline}
    < 0 ; \beta | A | 0 ; \beta > \; = \; \sum_{n}
    \sum_{m} < 0 ; \beta | n , \tilde{n} > < m , \tilde{m} | 0 ; \beta > < n ,
    \tilde{n} | A | m , \tilde{m} > \; = \\
     \sum_{n} \sum_{m} < 0 ; \beta | n , \tilde{n} > < m , \tilde{m} | 0 ; \beta
    > < n | A | n > \delta_{n ,m}
\end{multline}
where in the last passage we have used the equation \ref{eq:1th
decoupling condition}.

Hence:
\begin{equation}
    < 0 ; \beta | A | 0 ; \beta > \; = \; \sum_{n}  | < n , \tilde{n} | 0 ; \beta > |^{2} < n | A | n >
\end{equation}

So the equation \ref{eq:basic property of the thermal vacuum}
defining a \emph{thermal vacuum} is satisfied by the vector $ | 0
; \beta > \in \mathcal{H} \otimes \tilde{\mathcal{H}} $ if and
only if:
\begin{equation}
    | < n , \tilde{n} | 0 ; \beta > |^{2} \; = \; \frac{ \exp ( - \beta E_{n} ) }{Z(\beta)}
\end{equation}
that admits many solutions among which the simpler one may be
obtained imposing that $ < n , \tilde{n} | 0 ; \beta > \in
\mathbb{R} $:
\begin{equation} \label{eq:condition determining a thermal vacuum}
 < n , \tilde{n} | 0 ; \beta > \; := \; \frac{ \exp ( \frac{- \beta E_{n}}{2} )  }{\sqrt{Z( \beta)}}
\end{equation}

\bigskip

\begin{remark}
\end{remark}
Up to this point the introduction of the notion of a \emph{thermal
vacuum} may appear an unjustified complication.

Its power appears as soon as one analyzes the phenomenon of
\emph{symmetry breaking} and \emph{symmetry restoration} at
strictly positive temperature.

Let us, first of all, review the definition of symmetry breaking
at zero temperature.

Let us suppose to have a strongly continuous one-parameter unitary
group $ U_{\alpha} := \exp ( - i \alpha Q ) $ that is a symmetry
of the system, i.e.:
\begin{equation}
 U_{\alpha} H U_{\alpha}^{\dag} \; = \; H \; \; \forall \alpha \in
 \mathbb{R}
\end{equation}
and hence:
\begin{equation}
    [ Q , H ] \; = \; 0
\end{equation}

We will say that such a symmetry is \emph{broken at zero
temperature} (i.e. at $ \beta = + \infty $) whether:
\begin{equation}
    Q | 0 > \; \neq \; 0
\end{equation}
where $ | 0 > $ is the vacuum state.

 Let us define a \emph{Goldstone operator at zero temperature} (i.e. at $ \beta = + \infty $) as an operator A
 such that:
 \begin{equation}
    < 0 | [ Q , A ] | 0 > \; \neq \; 0
\end{equation}
Clearly the symmetry is \emph{broken at zero temperature} if and
only if there exists a \emph{Goldstone operator at zero
temperature}.

\smallskip

Let us now consider the same system in thermodynamical equilibrium
with a thermal bath at strictly positive temperature.

We will say that the symmetry is \emph{broken at inverse
temperature $ \beta \in [ 0 , + \infty ) $} whether:
\begin{equation}
    Q | 0 ; \beta > \; \neq \; 0
\end{equation}
where $ | 0 ; \beta > $ is a \emph{thermal vacuum}.

 Let us define a \emph{Goldstone operator at inverse temperature $ \beta \in [ 0 , + \infty ) $ } as an operator A
 such that:
 \begin{equation}
    < 0 ; \beta | [ Q , A ] | 0 ; \beta > \; \neq \; 0
\end{equation}
Clearly the symmetry is \emph{broken at inverse temperature $
\beta \in [ 0 , + \infty ) $} if and only if there exists a
\emph{Goldstone operator at inverse temperature $ \beta $}.

\newpage
\section{Expectation value of the bosonic phase operator}

Given the bosonic oscillator with hamiltonian given by the
equation \ref{eq:hamiltonian of the bosonic oscillator} the
condition of equation \ref{eq:condition determining a thermal
vacuum} determines the following \emph{thermal vacuum}:
\begin{equation} \label{eq:thermal vacuum for the bosonic oscillator}
    | 0 ; \beta > \; := \; \sqrt{1 - \exp( - \beta \omega )  }
    \sum_{n=0}^{\infty} \exp ( - \frac{n \beta \omega }{2} )| n ,
    \tilde{n} >
\end{equation}
where we have used the fact that:
\begin{equation}
    \sum_{n=0}^{\infty} x^{n} \; = \; \frac{1}{1-x} \; \; \forall
    x \in [0,1)
\end{equation}

Introduced the self-adjoint operator:
\begin{equation}
    Q( \phi_{B} ) \; := \; - i \phi_{B}( \beta ) ( \tilde{a} a - a^{\dag}
    \tilde{a}^{\dag})
\end{equation}
and the unitary operator:
\begin{equation}
 U( \phi_{B}  ) \; := \; \exp ( -i  Q( \phi_{B} ))
\end{equation}
it follows that the \emph{thermal vacuum} may be obtained by the
\emph{Bogoliubov transformation}:
\begin{equation}
    | 0 ; \beta > \; = \; U(\phi_{B}) | 0 , \tilde{0} >
\end{equation}
provided:
\begin{equation} \label{eq:1th condition on the parameter of the bosonic Bogoliubov transformation}
    \cosh \phi_{B} ( \beta ) \; = \; \frac{1}{ \sqrt{1 - \exp( - \beta \omega )  }}
\end{equation}
\begin{equation} \label{eq:2th condition on the parameter of the bosonic Bogoliubov transformation}
    \sinh  \phi_{B} ( \beta ) \; = \; \frac{\exp ( - \frac{ \beta \omega }{2}) }{ \sqrt{1 - \exp( - \beta \omega )  }}
\end{equation}

\smallskip

Clearly:

\begin{equation}
    < N_{B} >_{\beta} \; = \; < 0 ; \beta | N_{B} |  0 ; \beta > \; =
    \; [ 1 - \exp ( - \beta \omega )]  \, \sum_{n=0}^{\infty}  n \exp ( - \beta \omega n )  \; = \; \frac{ \exp ( - \beta \omega )  }{ 1 - \exp ( - \beta \omega )
    } \; = \; \sinh^{2} \phi_{B} ( \beta )
\end{equation}
where we have used the fact that:
\begin{equation}
  \sum_{n=0}^{\infty}  n x^{n}  \; = \; \frac{x}{ ( 1 - x)^{2}}
  \; \; \forall x \in [ 0 ,1 )
\end{equation}
Furthermore:
\begin{multline}
    < \exp ( i \hat{\theta} ) >_{\beta} \; = \;  < 0 ; \beta | \exp ( i \hat{\theta} ) |  0 ; \beta > \;
  =  \; ( 1 - \exp ( - \beta \omega ) ) \sum_{n=0}^{\infty}
  \sum_{k=0}^{\infty} \sum_{m=0}^{\infty} \exp [ - \frac{ (n+m) \beta \omega
  }{2} ] < n , \tilde{n} | k > < k +1 | m , \tilde{m} > \; = \\
 \sum_{n=0}^{\infty}
  \sum_{k=0}^{\infty} \sum_{m=0}^{\infty} \exp [ - \frac{ (n+m) \beta \omega
  }{2} ] \delta_{n,k} \delta_{m,k+1} \delta_{n,m} \; = \; 0
\end{multline}
as it can be  checked  computing the expectation value of the
\emph{bosonic exponential phase operator} directly, i.e. avoiding
the thermofield dynamics' approach:
\begin{multline}
     < \exp ( i \hat{\theta} ) >_{\beta} \; = \; \frac{ \sum_{n=0}^{\infty} \exp ( - \beta E_{n} ) < n |  \exp ( i \hat{\theta} ) | n >
     }{Z(\beta)} \; = \\
  ( 1 - \exp ( - \beta \omega ) ) \sum_{n=0}^{\infty}
  \sum_{k=0}^{\infty}  \exp ( - \beta \omega n ) \delta_{n,k}
  \delta_{n,k+1} \; = \; 0
\end{multline}

\newpage
\section{Expectation value of the fermionic phase operator} Given
the fermionic oscillator with hamiltonian given by the equation
\ref{eq:hamiltonian of the fermionic oscillator}, the condition of
equation \ref{eq:condition determining a thermal vacuum}
determines the following \emph{thermal vacuum}:
\begin{equation} \label{eq:thermal vacuum for the fermionic oscillator}
    | 0 ; \beta > \; := \; \frac{1}{ \sqrt{1 + \exp( - \beta \omega )  }
    } ( | 0 , \tilde{0} > + \exp ( - \frac{\beta \omega}{2} ) | 1 ,
    \tilde{1} > )
\end{equation}
Introduced the self-adjoint operator:
\begin{equation}
    Q( \phi_{F} ) \; := \; - i \phi_{F}( \beta ) ( \tilde{a} a - a^{\dag}
    \tilde{a}^{\dag})
\end{equation}
and the unitary operator:
\begin{equation}
 U( \phi_{F}  ) \; := \; \exp ( -i  Q( \phi_{F} ))
\end{equation}
it follows that the \emph{thermal vacuum} may be obtained by the
\emph{Bogoliubov transformation}:
\begin{equation}
    | 0 ; \beta > \; = \; U(\phi_{F}) | 0 , \tilde{0} >
\end{equation}
provided:
\begin{equation} \label{eq:1th condition on the parameter of the fermionic Bogoliubov transformation}
    \cos \phi_{F} ( \beta ) \; = \; \frac{1}{ \sqrt{1 + \exp( - \beta \omega )  }}
\end{equation}
\begin{equation} \label{eq:2th condition on the parameter of the fermionic Bogoliubov transformation}
    \sin  \phi_{F} ( \beta ) \; = \; \frac{\exp ( - \frac{ \beta \omega }{2}) }{ \sqrt{1 + \exp( - \beta \omega )  }}
\end{equation}

Clearly:
\begin{multline}
    < N_{F} >_{\beta} \; = \; < 0 ; \beta | N_{F} |  0 ; \beta > \; =
    \; \frac{1}{1 + \exp ( - \beta \omega )} [ < 0 , \tilde{0} | + \exp( - \frac{ \beta \omega  }{2}  ) < 1 , \tilde{1} | ] N_{F} [ | 0 , \tilde{0} > +  \exp( - \frac{ \beta \omega  }{2}  )  | 1 ,
    \tilde{1} >  ] \; = \\
\frac{1}{1 + \exp ( - \beta \omega )} [ < 0 , \tilde{0} | + \exp(
- \frac{ \beta \omega  }{2}  ) < 1 , \tilde{1} | ] \exp( - \frac{
\beta \omega  }{2}) | 1 ,
    \tilde{1} > \; = \\
     \frac{ \exp ( - \beta \omega )  }{ 1 + \exp ( - \beta \omega )
    } \; = \; \sin^{2} \phi_{F} ( \beta )
\end{multline}

\begin{equation}
    < \exp ( i \hat{\theta} ) >_{\beta} \;  = \; < 0 ; \beta | \exp ( i \hat{\theta} ) |  0 ; \beta > \;
    = \; \frac{ \exp ( - \beta \omega ) < 1 , \tilde{1} | 1 , \tilde{1} > }{ 1 + \exp ( - \beta \omega )
    } \; = \; \frac{ \exp ( - \beta \omega )  }{ 1 + \exp ( - \beta \omega )
    } \; = \; \sin^{2} \phi_{F} ( \beta )
\end{equation}

\newpage
\section{The supersymmetric phase operator as a Goldstone
operator at strictly positive temperature}
Given the
supersymmetric oscillator with hamiltonian given by the equation
\ref{eq:hamiltonian of the supersymmetric oscillator}, let us
introduce  the self-adjoint operator:
\begin{equation}
    G( \phi_{B} ,  \phi_{F}  ) \; :=  \; - i \phi_{B}( \beta ) ( \tilde{a}_{B} a_{B} - a_{B}^{\dag}
    \tilde{a}_{B}^{\dag}) - i \phi_{F}( \beta ) ( \tilde{a}_{F} a_{F} - a_{F}^{\dag}
    \tilde{a}_{F}^{\dag})
\end{equation}
and the unitary operator:
\begin{equation}
 U( \phi_{B},  \phi_{F}  ) \; := \; \exp ( -i  G( \phi_{B} , \phi_{F} ))
\end{equation}

The \emph{thermal vacuum} determined by the equation
\ref{eq:condition determining a thermal vacuum} may be obtained by
the \emph{Bogoliubov transformation}:
\begin{equation}
    | 0 ; \beta > \; = \; U( \phi_{B},  \phi_{F}  ) | 0 , \tilde{0} >
\end{equation}
provided:
\begin{equation} \label{eq:condition on the parameter of the supersymmetric Bogoliubov transformation}
    \tanh \phi_{B} ( \beta ) \; = \;  \tan \phi_{F} ( \beta ) \; =
    \; \exp( - \frac{ \beta \omega }{2} )
\end{equation}

\smallskip

Clearly:

\begin{equation}
    < N_{B} >_{\beta} \; = \; < 0 ; \beta | N_{B} |  0 ; \beta > \; =
    \;  \sinh^{2} \phi_{B} ( \beta )
\end{equation}
\begin{equation}
     < N_{F} >_{\beta} \; = \; < 0 ; \beta | N_{F} |  0 ; \beta > \; =
    \;  \sin^{2} \phi_{F} ( \beta )
\end{equation}

so that the \emph{internal energy} is:
\begin{equation}
    \mathcal{U} ( \beta ) \; = \; < 0 ; \beta | H | 0 ; \beta > \;
    = \omega [ sinh^{2}  \phi_{B} ( \beta ) +  \sin^{2} \phi_{F} ( \beta
    )]
\end{equation}

\smallskip

Furthermore:
\begin{equation}
    Q | 0 ; \beta > \; = \; a_{B}^{\dagger} a_{F} | 0 ; \beta > \;= \;  \; \cosh \phi_{B} ( \beta ) \sin \phi_{F}
    ( \beta ) | n_{B}( \beta ) = 1 , n_{F} ( \beta ) = 0 ;
    \tilde{n}_{B} ( \beta ) = 0 , \tilde{n}_{F} ( \beta ) = 1 >
\end{equation}
\begin{equation}
    \bar{Q} | 0 ; \beta > \; = \; a_{F}^{\dagger} a_{B} | 0 ; \beta > \;= \;  \; \sinh \phi_{B} ( \beta ) \cos \phi_{F}
    ( \beta ) | n_{B}( \beta ) = 0 , n_{F} ( \beta ) = 1 ;
    \tilde{n}_{B} ( \beta ) = 1 , \tilde{n}_{F} ( \beta ) = 0 >
\end{equation}
and hence:
\begin{equation}
    Q  | 0 ; \beta > \; \left\{%
\begin{array}{ll}
    = 0, & \hbox{if $ \beta = + \infty $;} \\
    \neq 0, & \hbox{if $ \beta \in ( 0 , + \infty $).} \\
\end{array}%
\right.
\end{equation}
\begin{equation}
    \overline{Q}  | 0 ; \beta > \; \left\{%
\begin{array}{ll}
    = 0, & \hbox{if $ \beta = + \infty $;} \\
    \neq 0, & \hbox{if $ \beta \in ( 0 , + \infty $).} \\
\end{array}%
\right.
\end{equation}
from which we can infer that:
\begin{itemize}
    \item the supersymmetry is unbroken  at zero temperature
    \item the supersymmetry is broken at every
    temperature $ T > 0$ (finite or infinite).
\end{itemize}

\bigskip

\begin{remark}
\end{remark}
Supersymmetry breaking is usually analyzed in terms of the
\emph{Witten index} (defined as the difference between the number
of bosonic and fermionic zero-energy states).

Indeed, in his 1982's fundamental paper, Edward Witten showed that
the vanishing of the \emph{Witten index} is a necessary (though
not sufficient) condition for having Susy broken.

Unfortunately, in the case of the supersymmetric oscillator, the
computation of Witten index involves subtle regularization's
issues that we have preferred to avoid (see for instance the $
4^{th} $ chapter "SUSY Breaking, Witten Index and Index Condition"
of \cite{Bagchi-01} and the references therein indicated).

\bigskip

We will now show that the \emph{supersymmetric phase operator} is
a \emph{Goldstone operator} at every temperature $ T > 0$.

Let us observe, first of all, that:
\begin{equation}
    Z( \beta ) \; = \; \sum_{n_{B}=0}^{ \infty}
    \sum_{n_{F}=0}^{1} \exp [ - \beta \omega ( n_{B} + n_{F} ) ]
    \; = \;  \sum_{n_{B}=0}^{+ \infty} \exp ( - \beta \omega
    n_{B} ) + \exp [ - \beta \omega
    (n_{B} +1 )] \; = \; \frac{1 + \exp ( - \beta \omega ) }{1 - \exp ( - \beta \omega ) }
\end{equation}
Furthermore some trivial computation leads to:
\begin{equation}
    < n_{B} , n_{F} | [ Q , \exp ( i \hat{\theta} ) ] | n_{B} ,
    n_{F} > \; = \; \exp ( i \theta ) \sum_{n=0}^{\infty}
    \sqrt{n+1} \delta_{n_{B},n+1} \delta_{n_{F},0}
\end{equation}
Therefore:
\begin{multline} \label{eq:supersymmetric phase as a Goldstone operator1}
    < 0 ; \beta | [ Q , \exp ( i \hat{\theta} ) ] | 0 ; \beta > \;
     = \frac{Tr \exp ( - \beta H ) [ Q , \exp ( i \hat{\theta} )
     ] }{Z(\beta)}
     \; = \\
     \frac{ \sum_{n_{B}=0}^{ \infty}
    \sum_{n_{F}=0}^{1} \exp [ - \beta \omega ( n_{B} + n_{F} ) ]  < n_{B} , n_{F} | [ Q , \exp ( i \hat{\theta} ) ] | n_{B} ,
    n_{F} > }{Z( \beta
    )} \; = \\
     \frac{1 - \exp ( - \beta \omega ) }{1 + \exp ( - \beta \omega ) } \exp ( i \theta ) Li_{-\frac{1}{2}} [ \exp ( -
    \beta \omega ) ] \; \neq \; 0 \; \; \forall \beta \in [ 0 , +
    \infty )
\end{multline}
where:
\begin{equation}
    Li_{n} ( x) \; := \; \sum_{k=1}^{\infty} \frac{ x^{k} }{k^{n} }
\end{equation}
is the \emph{polylogarithmic function.}

In a similar way one gets that:
\begin{equation}
    < n_{B} , n_{F} | [ \bar{Q} , \exp ( i \hat{\theta} ) ] | n_{B} ,
    n_{F} > \; = \; -  \exp ( i \theta ) \sum_{n=0}^{\infty}
    \sqrt{n+1} \delta_{n_{B},n} \delta_{n_{F},1}
\end{equation}
and hence:
\begin{multline}  \label{eq:supersymmetric phase as a Goldstone operator2}
    < 0 ; \beta | [ \bar{Q} , \exp ( i \hat{\theta} ) ] | 0 ; \beta > \;
     = \frac{Tr \exp ( - \beta H) [ \bar{Q} , \exp ( i \hat{\theta} )
     ]}{ Z(\beta)  } \; = \\
      \frac{ \sum_{n_{B}=0}^{ \infty}
    \sum_{n_{F}=0}^{1} \exp [ - \beta \omega ( n_{B} + n_{F} ) ]  < n_{B} , n_{F} | [ Q , \exp ( i \hat{\theta} ) ] | n_{B} ,
    n_{F} > }{Z( \beta
    )} \; = \\
    - \frac{1 - \exp ( - \beta \omega ) }{1 + \exp ( - \beta \omega ) } \exp ( i \theta ) Li_{-\frac{1}{2}} [ \exp ( -
    \beta \omega ) ] \; \neq \; 0 \; \; \forall \beta \in [ 0 , +
    \infty )
\end{multline}

\bigskip

\begin{remark}
\end{remark}
Let us remark that the equation \ref{eq:supersymmetric phase as a
Goldstone operator1} and the equation \ref{eq:supersymmetric phase
as a Goldstone operator2} contemplate also the case $ \beta = 0 $
corresponding to infinite temperature.

In fact, it can be easily checked that, in the limit $ \beta
\rightarrow 0 $, the divergence of $ Li_{-\frac{1}{2}} [ \exp ( -
    \beta \omega ) ] $ wins against the convergence to zero of $ \frac{1 - \exp ( - \beta \omega ) }{1 + \exp ( - \beta \omega )} $.

Therefore the \emph{supersymmetric phase operator} is a Goldstone
operator at infinite temperature.

\bigskip

\begin{remark}
\end{remark}
Let us remark that since
\begin{equation}
    \lim_{\beta \rightarrow + \infty} < 0 ; \beta | [ Q , \exp ( i \hat{\theta} ) ] | 0 ; \beta
    > \; = \; \lim_{\beta \rightarrow + \infty} < 0 ; \beta | [ \bar{Q} , \exp ( i \hat{\theta} ) ] | 0 ; \beta
    > \; = \; 0
\end{equation}
it follows that the \emph{supersymmetric phase operator} is not a
Goldstone operator at zero temperature, as we already knew by the
fact that the supersymmetry is unbroken at zero temperature.

\end{document}